\documentclass[a4paper]{jpconf}
\usepackage{graphicx}
\usepackage{citesort}

\newcommand{\eq}{\begin{eqnarray}}
\newcommand{\en}{\end{eqnarray}}

\newcommand{\beq}{\begin{equation}}
\newcommand{\eeq}{\end{equation}}
\newcommand{\ba}[1]{\begin{eqnarray} \label{(#1)}}
\newcommand{\ea}{\end{eqnarray}}

\newcommand{\barr}{\begin{eqnarray}}
 \newcommand{\earr}{\end{eqnarray}} 
\begin{document}
\title{Neutrino oscillometry}

\author{J.D.Vergados$^1$, Y. Giomataris$^2$ and Yu. N. Novikov$^{3}$}

\address{1 Physics Department, University of Ioannina, Ioannina, Greece}
\ead{vergados@uoi.gr}
\address{
2 CEA, Saclay, DAPNIA, Gif-sur-Yvette, Cedex,France}
\address{
3 Petersburg Nuclear Physics Institute, 188300, Gatchina,Russia}

\begin{abstract}
Neutrino oscillations are studied employing sources of low energy monoenergetic neutrinos 
following electron capture by the nucleus and measuring electron recoils. Since the neutrino energy is very low the oscillation length $L_{23}$
appearing in this electronic neutrino  disappearance experiment can be so small that the full oscillation can take place inside the detector so that one may determine very
accurately the neutrino oscillation parameters.
 In particular, since the oscillation probability is proportional to $\sin^2{2 \theta_{13}}$,  one can  measure or set a better limit on the unknown parameter $\theta_{13}$.  One, however, has to pay the price that  the expected counting rates are very small. Thus one needs a very intensive neutrino source and a large detector with as low as possible energy threshold and high energy and position resolution. Both spherical gaseous and cylindrical liquid detectors are studied. Different  source candidates are considered
\end{abstract}
\section{Introduction.}
The discovery of neutrino oscillations can be considered as one of the greatest triumphs of modern physics.
It began with atmospheric neutrino oscillations \cite{SUPERKAMIOKANDE}interpreted as
 $\nu_{\mu} \rightarrow \nu_{\tau}$ oscillations, as well as
 $\nu_e$ disappearance in solar neutrinos \cite{SOLAROSC}. These
 results have been recently confirmed by the KamLAND experiment \cite{KAMLAND},
 which exhibits evidence for reactor antineutrino disappearance.
  As a result of these experiments we have a pretty good idea of the neutrino
mixing matrix and the two independent quantities $\Delta m^2$, e.g $|m_2^2-m^2_1|$ and $|m^2_3-m^2_2|$.
 Fortunately these
two  $\Delta m^2$ values are vastly different, $$\Delta
m^2_{21}=|m_2^2-m_1^2|=(7.65^{+0.23}_{-0.20})\times 10^{-5}(eV)^2,\quad
\Delta m^2_{32}=|m_3^2-m_2^2|=(2.4^{+0.12}_{-0.11})\times 10^{-3}(eV)^2.$$
 This means that the relevant $L/E$ parameters are very different. Thus for a given energy the experimental results can approximately be described as two generation oscillations. For an accurate description  of the data, however, a three generation analysis  \cite{BAHCALL02},\cite{BARGER02} is necessary.

 In all of these analyses the
oscillation length is much larger than the size of the detector. So one is able to see the effect, if the detector is
placed in the right distance from the source. 

The most precise and unambiguous way to measure neutrino oscillations would 
be to determine changes in the flux of the given flavor of neutrinos over 
the entire oscillation length. Since the oscillation length is proportional 
to neutrino energy, the proper neutrino oscillometry would require a 
detector hundreds or even thousands of kilometers long if used with the 
present or proposed neutrino beams! As this is unrealistic, all beam 
experiments aiming at neutrino oscillations consider just a single or at 
most two point measurements instead of the full oscillometric approach. Also 
when using reactor neutrinos, the distance from the source to the first 
minimum is about 2 km - still beyond the current technological and 
financial boundaries for a detector. To be able to perform neutrino 
oscillometry using a realistic-size detector like LENA (100 m long) one 
needs a strong source of monoenergetic neutrinos with the energy of a few 
hundred of keV. Such a source could be produced in a nuclear reactor making 
neutrino oscillometry with LENA possible \cite{VerNov10}. Neutrino 
oscilometry provides a competitive and considerably less expensive 
alternative to long baseline neutrino beams.

The best way to detect low energy electron neutrinos  is by 
measuring electron recoils from neutrino-electron scattering. The total neutrino electron scattering cross section  be cast in the form:
\beq
\sigma(L,x,y_{\tiny{th}})=\sigma(0,x,y_{\tiny{th}})\left( 1-\chi(x,y_{\tiny{th}}) p(L,x)\right )
\label{sigmatot1}
\eeq
with $x=\frac{E_{\nu}}{m_e}$ and $y_{\tiny{th}}=\frac{(T_e)_{\tiny{th}}}{m_e}$, with $(T_e)_{\tiny{th}}$ the threshold electron energy imposed by the detector and 
\beq
p(L,x)=\sin ^2\left(\frac{ 0.595922 L}{33x}\right) \sin ^2(2 \theta_{solar} )+
\sin^2\left(\frac{ 0.595922L}{x}\right) \sin ^2\left(2 \theta_{13}\right) 
 \eeq
 with $L$ the source detector distance in meters. The functions  $ \sigma(0,x,y_{\tiny{th}})$, the cross section in the absence of oscillation, and $\chi(x,y_{\tiny{th}}) $, which takes care of the other neutrino flavors, have been previously described \cite{VerNov10}. The oscillation length of interest to us take the form:
\beq
 L_{32} =  \frac{2.48[\mbox{m}] E_{\nu}}{\Delta m^2_{32}([\mbox{eV}])^2}
  \Rightarrow
 L_{32}[\mbox{m}]\approx E_{\nu} [\mbox{keV}]
 \label{L32}
\eeq                                                             
 The values in the square brackets in Eqs (\ref{L32}) indicate the dimensions used.
 
\begin{figure}[!ht]
 \begin{center}
 % \rotatebox{90}{\hspace{0.0cm} {$R_0\frac{dN}{dL}\longrightarrow\Lambda=130$y$^{-1}$}}
\includegraphics[width=2.3in,height=2.0in]{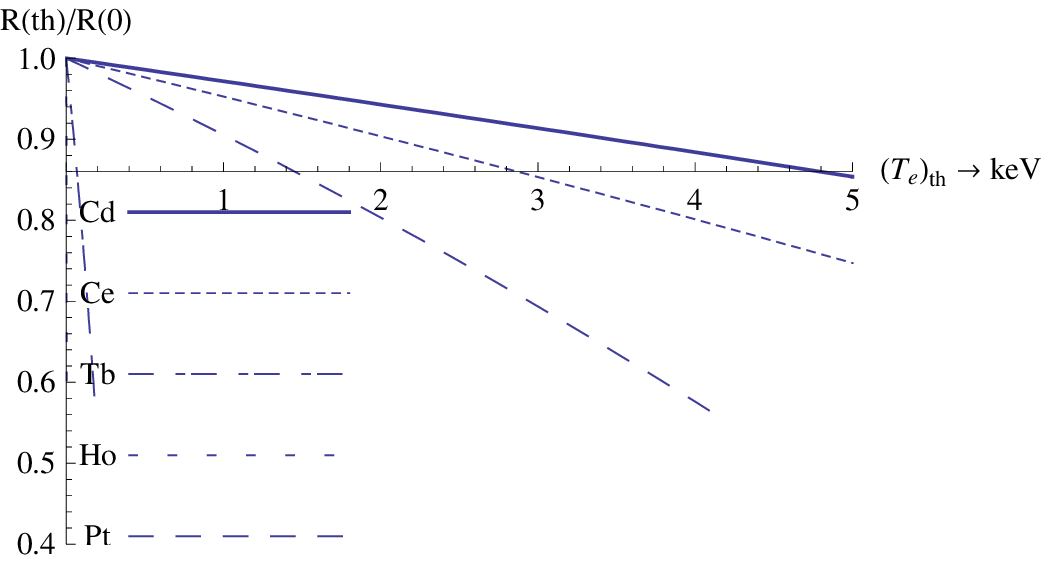}
\includegraphics[width=2.3in,height=2.0in]{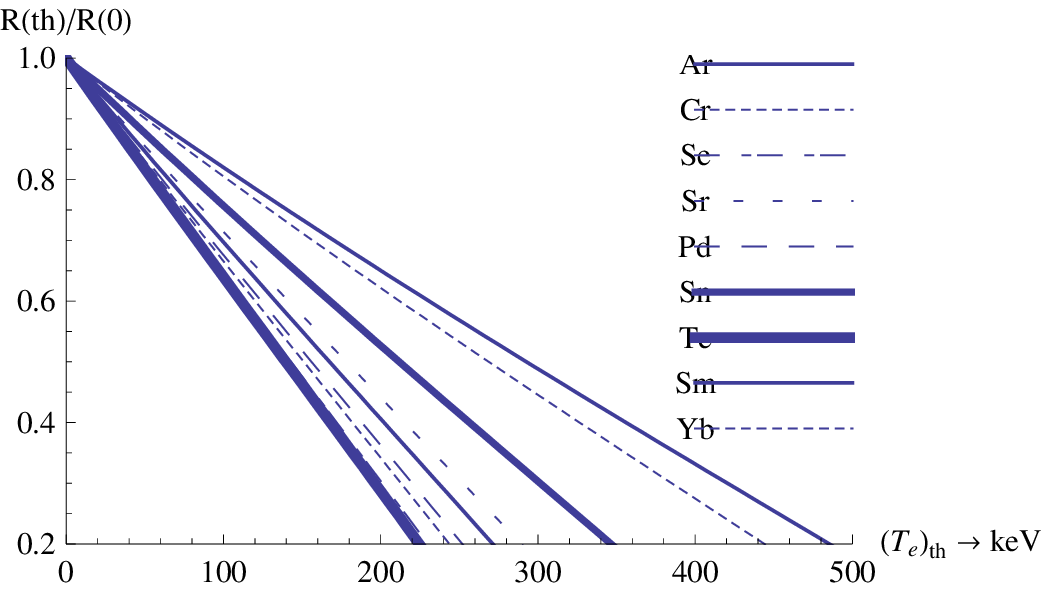}
%\hspace*{-0.0cm} { $E_{\text{th}} \rightarrow$keV}\\
 \caption{The dependence of the rate on the energy threshold, $T_{{th}}$, in the case of a gaseous spherical TPC detector on the left and the LENA detector on the right.}
 \label{spcylth}
  \end{center}
  \end{figure}

%As can be seen from Table \ref{tab1} (column 5) one can roughly divide the 
The neutrino sources of interest are divided into two categories: Those which have $L_{32}\le 50$ m and those with $L_{32} > 110$ m. 
For the former nuclides the TPC counting method can be used in the gas-filled NOSTOS sphere approach \cite{NOSTOS1}, \cite{Giomataris}, whereas for both and mainly for the latter category 
with the larger $L_{32}$, the long liquid scintillator (LS) detector \cite{OBERAUER} is  preferable. One of the main advantages of the spherical TPC detectors is the very low energy threshold \cite{LETHR10} they can achieve (0.1 keV), which allows them to take advantage of the very low energy neutrinos. From this point of view a comparzon between the two types of detectors is given in Fig. \ref{spcylth}.
  
\begin{figure}[!ht]
 \begin{center}
 \rotatebox{90}{\hspace{2.0cm} {$\frac{dN}{dL}\rightarrow$ m$^{-1}$}}
 \includegraphics[scale=1.0]{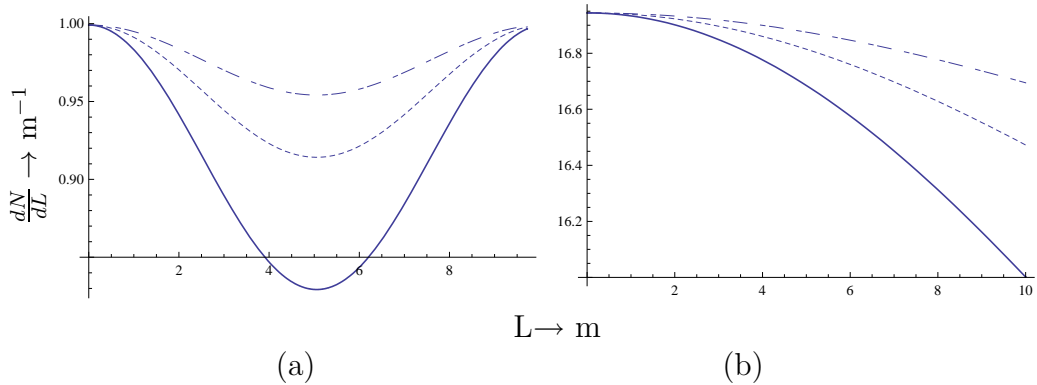}
 \caption{The  rate $\frac{dN}{dL}$ (per meter) for Ar at $10$ Atm with $1$ Kg of
 $^{157}$Tb (a) and $^{193}$Pt (b)
 as a function of
 the source-detector distance (in m ).
%At the bottom we plot all the cases in the same plot.
The results shown correspond to
$\sin^2{2\theta_{13}}=0.170,0.085$ and $0.045$ (decreasing from bottom to top). This rate was obtained for a running period equal to the half life of the source.  The analysis is much simpler than that for the cylindrical geometry since the geometric factor $g_{av}$ for the spherical detector is unity.
%This rate must be multiplied by $1-e^{-t/\tau}$ to get the number of events afterrunning time $t$.
}
 %\end{center}
 \label{rates1and4_40}
  \end{center}
  \end{figure}

For a spherical detector two typical examples, obtained with a threshold of 0.1keV,  are shown in Fig. \ref{rates1and4_40}. Clearly a compromise has to be made to achieve   as large as possible portion of the oscillation inside the detector with a reasonable detection rate.
Since the beautiful results of the first category  have  been previously discussed \cite{NOSTOS1}, \cite{VerNov10}, in this paper we will concentrate on the second category. 
\section{Short baseline neutrino oscillations-The detection principle}

 Any  change of scattering rate as a function of position, in excess of the geometric factor would give, for the first time, a continuous 
(oscillometric) measure of flavor disappearance. 

The detector for $\nu-e$ scattering events should be as long as 
possible and should have a large fiducial volume. It's energy registration 
threshold should be as law as possible since to have short oscillation 
length one needs low energy neutrinos. Presently only liquid scintillator 
(LS) technology can provide the required low detection threshold of 200 keV 
\cite{BOREXINO}. The proposed LENA detector would match these requirements. 
Moreover, with the suggested length of about 100 m LENA would be the longest 
LS detector ever build. Due to the improved signal processing and timing 
characteristics as compared to the operating LS detectors \cite{BOREXINO} the 
expected position sensitivity of LENA will be better than 50 cm even at the 
level of a few hundreds of keV of the recoil energy of the electron. The 
energy resolution would be $\sim $10 {\%} in this energy region \cite{OBERAUER}.

As the cross-sections for $\nu-e$ scattering are tiny, a very strong neutrino source should be used to provide adequate statistics. 
Fortunately, there are many nuclei decaying via electron capture (EC). Since 
EC is a two body process the emitted electron neutrino is monoenergetic and 
carries most of the transition energy. Table \ref{tab:sources} lists some of the isotopes 
decaying via EC with suitable $Q$ values to produce monoenergetic neutrinos of 
a few hundreds of keV and with half-lifes of a few months allowing for 
convenient handling. 
They are relatively easy to produce via 
neutron capture reaction, see, e.g., the GALLEX experiment \cite{GALLEX} for the 62 PBq $^{51}$Cr source. 

\begin{table}[htbp]
\begin{center}
\caption{ Neutrino sources which could be produced by irradiation in the 
neutron reactors. The intensities of neutrino sources per second have been 
estimated per 1 kg of the target element with the natural isotope abundances 
and assuming a 10 day irradiation with the neutron flux of $5\times10^{14}$ 
n/cm$^{2}$/s. Neutron capture cross sections were taken from 
[http://ie.lbl.gov/].}
\label{tab:sources}
%\begin{tabular}{|l|p{36pt}|l|l|l|p{73pt}|p{76pt}|}
\begin{tabular}{|c|c|c|c|c|c|c|}
\hline
Nuclide& 
$T_{1/2,}$& 
$Q_{\varepsilon }$(keV)& 
$E_{\nu }$& 
$E_{e,max}$& 
Ir. target & 
$\nu$-intensity  \\
& d& (keV)& (keV)&(keV) &(10 d) &(s$^{-1}$) \par(per kg)\\
%\hline
%Nuclide& 
%T$_{1/2,}$ \par days& 
%Q$_{\varepsilon }$(keV)& 
%E$_{\upsilon }$=Q$_{\varepsilon }$-B$_{i}$---E$_{\gamma }$(keV)& 
%E$_{e,max}$(keV)& 
%Irradiated target during 10 days \par & 
%õ-intensity \par (per 1 kg of target, per s) \\
\hline
$^{37}$Ar& 
35 & 
814& 
811 (100{\%})& 
617& 
Ar& 
8.3x10$^{15}$ \\
\hline
$^{51}$Cr& 
28 & 
753& 
747 (90{\%})& 
560& 
$^{50}$Cr& 
2.3x10$^{16}$ \\
\hline
$^{75}$Se& 
120 & 
863& 
450 (96{\%})& 
287& 
Se& 
1.1x10$^{14}$ \\
\hline
$^{113}$Sn& 
116 & 
1037& 
617 (98{\%})& 
436& 
Sn& 
8x10$^{11}$ \\
\hline
$^{145}$Sm& 
340 & 
616& 
510 (91{\%})& 
340& 
Sm& 
2x10$^{12}$ \\
\hline
$^{169}$Yb& 
32 & 
910& 
470 (83{\%})& 
304& 
Yb& 
1.1x10$^{15}$ \\
\hline
\end{tabular}
\end{center}
\end{table}

%\section{Short baseline neutrino oscillations}

The number of events in between $L$ and $L+dL$, where $L$ is the distance between the center of the source and the detection point, can be 
written in the following form \cite{VerNov10}:

 \beq
  R_0\frac{dN}{dL}=f_{\Phi} \Lambda g_{av}(u,L/R_0)\tilde{\sigma}(L,x,y_{\tiny{th}}),
  \label{Eq:gav}
\eeq
where 
\beq
\Lambda=\frac{G^2_F m^2_e}{2 \pi} R_0 N_{\nu} n_e 
\eeq
with $N_{\nu}$ the number of neutrinos emitted by the source, $n_e $ the density of electrons in the target ($n_{e} = 3\times 10^{29}$ m$^{-3}$ for LENA), $R_0$ the radius of the target and $\tilde{\sigma}(L,x,y_{\tiny{th}})$ is the neutrino - electron cross section in units of 
${G^2_F m^2_e}/{2 \pi}$. The quantity $f_{\Phi}$ reflects the fraction of the total flux relevant for the detector ($f_{\Phi}$=1, and 1/2 for a spherical detector (with the source at the center) and a cylindrical detector (with the source at the center of one of its bases respectively).  The geometric factor $g_{av}(u,L/R_0)$ in the case of a spherical detector is unity, while for a cylindrical geometry has been  previously given \cite{VerNov10}.

If one is content in extracting the value of the mixing angle only, this can be achieved by integrating the event rates over all $L$ in the detector. This essentially involves integrating the cross section over $L$, folded with the function $g_{{av}}(u,L/R_0) $, i.e. integrating   Eq. (\ref{Eq:gav}) over the $L$-values allowed by the detector. For sufficiently small mixing angle one can show that the event rate (in units of $\Lambda$) takes the form:
  \beq
  \frac{N}{\Lambda}=-A \sin^2{2 \theta_{13}}+B
  \eeq
  for $^{51}$Cr $A=0.048304$ and B=$ 0.982456$.
Then, depending on  the specifics of the experiment total number of events, $N_{0}$ can 
be presented in the form:
\beq
N_{0} = - a \sin^{2}{2\theta_{13}} + b
\label{totalevents}
\eeq 
%where $ a = 2.24\times 10^{5 }$ and $b = 4.55\times10^{6}$ for $^{51}$Cr and 275 days of 
%data handling. 
%It is seen from formula (\ref{totalevents}) that the total number of events should be 
%sensitive to the mixing angle $\theta_{13}$.

\section{The physics case for neutrino oscillometry}

Neutrino oscillometry offers an elegant way to solve a number of questions 
related to neutrino oscillations: a precise determination of the mixing 
angle $\theta_{13 }$and the oscillation length $L_{23}$, confirmation of the 
results of the ``global'' analysis of the oscillation data, and 
determination of the neutrino mass hierarchy. The latter would require a 
simultaneous long baseline measurement with the same detector. 
\subsection{Determination of the mixing angle $\theta_{13 }$}
%\underline {Mixing angle è}$_{13 }$

The big advantage of the short baseline oscillometry is that there is no 
matter influence in the observed events. As it is well known \cite{BARGER02},\cite{BarMag02} 
this matter effect gives degeneracy in the determination of the oscillation 
parameters in the long baseline experiments and should also be taken into 
consideration in some oscillation experiments with the reactor 
antineutrinos.

\begin{figure}[htbp]
\begin{center}
\includegraphics[width=3.13in,height=2.35in]{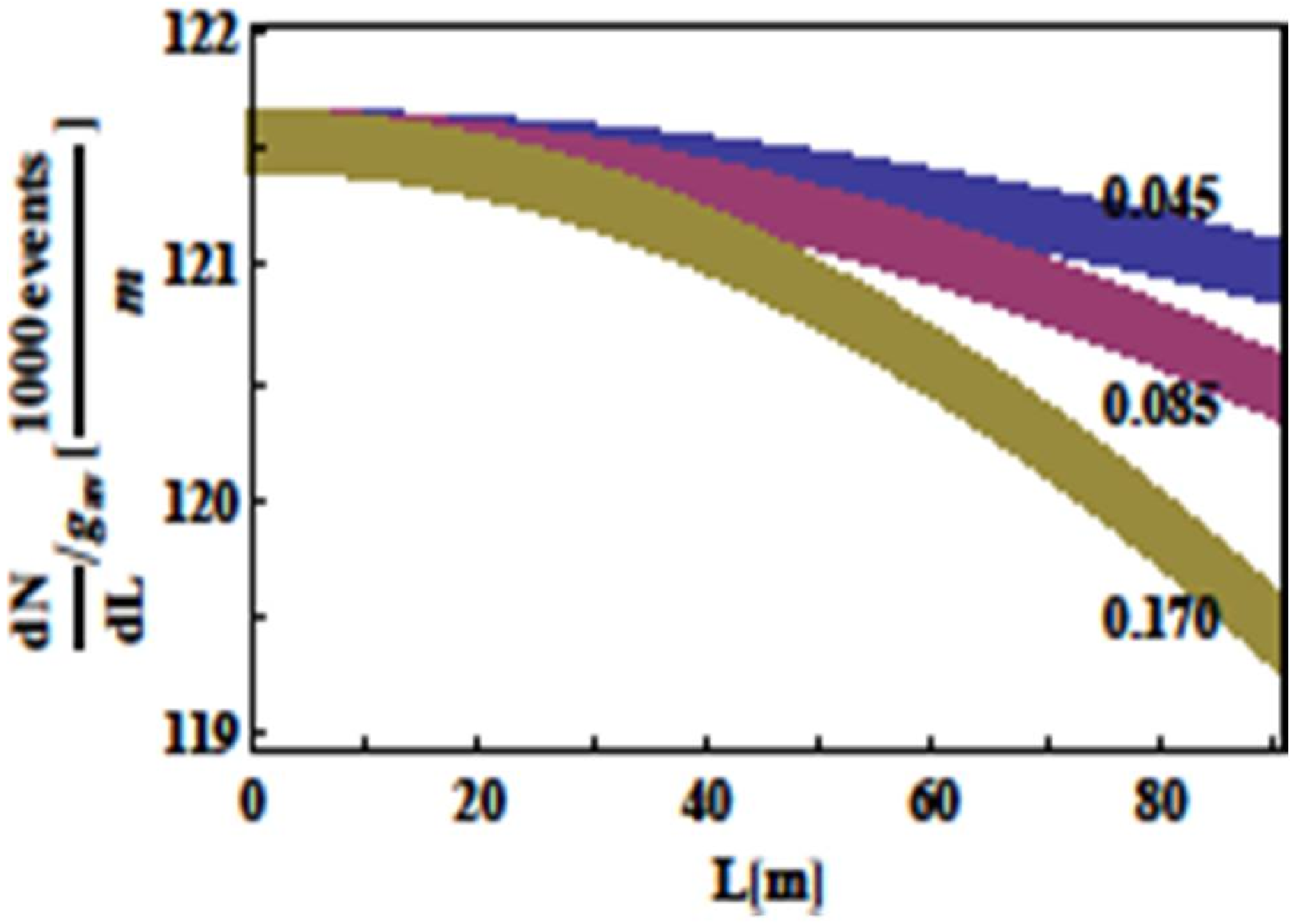}
\rotatebox{90}{\hspace{0.0cm} {$g_{{av}}(u,L/R_0)\rightarrow$}}
\includegraphics[width=2.13in,height=2.35in]{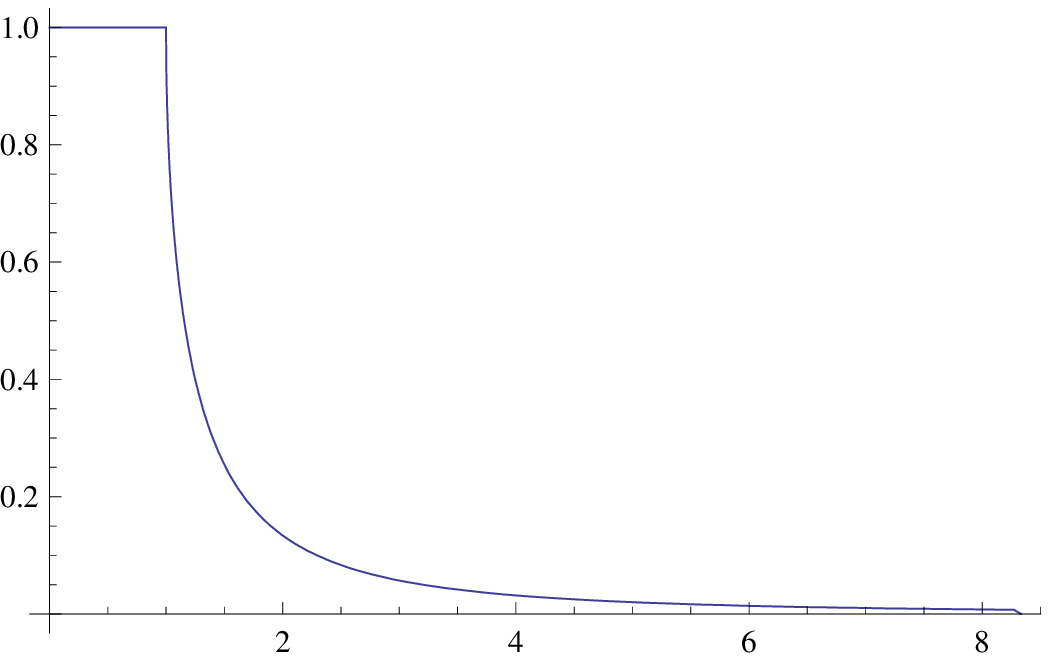}
\hspace*{-0.0cm} { $L/R\rightarrow$}
\caption{The number of the $\nu-e$ scattering events over the length $dL$ divided by the 
geometric factor $g_{av}$. The calculated values correspond to 5 data 
acquisition cycles, 55 days each with the $^{51}$Cr source installed at the 
top of LENA tank. The bottom (green), the middle (red) and the top (blue) 
correspond to $\sin^{2}2\theta_{13 }$= 0.170, 0.085 and 0.045, respectively. The geometric factor for $u=11/90$ is shown on the right.}
\label{fig1}
\end{center}
\end{figure}

The angle $\theta_{13 }$ can be determined from the analysis of both differential 
number of $\nu-e$ scattering events related to the length dL and the total 
number of events $N_{0}$ collected during the time of data acquisition in 
the full volume of LENA-detector. Differential curves for the neutrino 
scattering from the source $^{51}$Cr are shown in Fig. \ref{fig1}. As can be seen from 
this figure, the curves for different mixing angles $\sin^{2}{2\theta_{13 }}$ are 
well separated within the length of LENA detector. 

%The mixing angle can be determined from the total number of events \cite{LENAWP} via 
%equation (\ref{totalevents}). In Fig. \ref{fig2} we show the confidence level for the determination of 
%$\sin^{2}{2è_{13 }}$ as a function of its value calculated for two values of 
%the detection threshold (200 and 250 keV). The uncertainly of $\theta_{13 
%}$includes statistical errors, background events, uncertainty in neutrino 
%energy, energy registration threshold, acquisition time, and other, less 
%significant uncertainties. One can see that $\sin^{2}{2\theta_{13}}$ can be 
%obtained with 95{\%} confidence level ($2\sigma$) for $\sin{^{2}2\theta_{13}} \quad \approx $ 
%0.025. The calculations did not take into account the uncertainty in the $\nu-e $
%cross-section which is presently unknown for the energy region of $^{51}$Cr 
%neutrinos. Therefore the curves in Fig. \ref{fig2} serve only as a guideline. 
%Nevertheless the expected sensitivity of LENA is very competitive with those 
%of the other projects which are planned implemented by end of this decade 
%(Double CHOOZ: $\sin^{2}{2\theta_{13}} \quad \to $ 0.03; RENO $\to $ 0.02; Daya Bay 

%$\to $ 0.01; T2K $\to $ 0.01; and NOvA $\to $0.01). 
%Analysis of sensitivity from the differential curves could also be done \cite{LENAWP}
%

\subsection{The short oscillation length $L_{23}$.}

The value L$_{23 }$ can also be deduced from the oscillometry curves. This 
value can be compared with the neutrino energy which is usually well known, 
or can be measured independently very precisely \cite{BNW10}. For $^{51}$Cr 
the neutrino energy is presently known with the precision of 0.05 {\%}. 
Since  Eq. (\ref{L32}) is valid, if the value of global analysis for 
$\Delta m_{23}^{2 }= 2.5\times 10^{-3}$ (eV)$^{2}$ is used, this comparison will be 
helpful for assessment of the global analysis itself.

\subsection{Neutrino mass hierarchy. }

As the oscillometry method provides precise determination of $\theta_{13 }$ free 
of 8-fold degeneracy, the long baseline measurements in the same detector 
(LENA) with ì-neutrino beam would yield the information on the neutrino mass 
hierarchy without the need to change neither the energy of the neutrino beam 
nor the detector position. In this case the CERN- Pyhas$\ddot{a}$lmi combination \cite{PelSar06} 
 looks quite promising for such type of measurements. Since 
short and long baseline experiments are disentangled by the energy region, 
both $\theta_{13}$ and the sign of ${\{}\Delta m_{13}^{2}${\}} experiments can be implemented 
using the same detector -- LENA in Pyhas$\ddot{a}$lmi.

%\begin{figure}[htbp]
%\centerline{\includegraphics[width=4.46in,height=3.66in]{fig2.eps}}
%\caption{The dependence of the confidence level determined from the total number 
%of $\nu-e$ scattering events in LENA-detector on the expected value of the 
%mixing angle. The calculations were made for two values of the lower level 
%registration thresholds $T_{th}$=200 keV (the upper curve) and 250 keV 
%(the lower curve). The calculated values correspond to 5 data acquisition 
%cycles, 55 days each with the $^{51}$Cr source installed at the top of LENA 
%tank.}
%\label{fig2}
%\end{figure}

\section{Conclusions}

We have discussed the importance of neutrino oscillometry involving low energy monocromatic neutrinos. Ideally one would like to employ gaseous TPC detectors with an extremely low energy threshold of 0.1 keV and neutrino sources with energy less than 50 keV. At present, however, one may have to content with a compromise, i.e. employ liquid detectors and use neutrino sources with  energy of a few hundreds of keV. To this end the 
LENA detector is exceptionally well suited to perform precise determination of 
neutrino oscillation parameters thanks to the relatively low detection threshold ($\sim 
$200 keV) and considerable length ($\sim $100m). The needed electron-capture 
source emitting high-intensity monoenergetic and low-energy neutrinos can be 
manufactured by neutron irradiation in the core of a reactor. The 
disappearance of electron neutrinos can be followed over the full length of 
the detector by registering neutrino-electron scattering events. The 
resulting oscillometric curve and the total number of the events will 
provide accurate determination of the mixing angle $\theta_{13}$.
% For example, 
%using the $^{51}$Cr source, LENA would determine $\sin^{2}{2\theta_{13 }}<$ 0.03 
%with a 95 {\%} CL already within a couple of years of operation. A longer 
%run would improve this value further.
 The main advantages of the gaseous TPC detectors are:
 \begin{itemize}
 \item The energy threshold can be very low.
 \item One can explore real low energy neutrinos.
 \item The geometry is simple. The only L-dependence of the event rate comes from the oscillation.
 \end{itemize}
 The disadvantage is that, for at present realistic neutrino sources, the event rate is small. Furthermore the solar neutrino background may be serious. It does not, however, depend on $L$ and, if necessary, it  can be measured.
 
 The main advantages of neutrino oscillometry with LENA are 
summarized as follows:
\begin{itemize}
\item The short oscillation length $L_{23}$ can be determined directly and the value of $\theta_{13 }$
%can be determined
 very precisely, without being affected by the 8-folded degeneracy. 
\item The mass hierarchy can be measured simultaneously with the same detector by performing a long baseline experiment (preferably CERN-Pyhas$\ddot{a}$lmi) and using the determined $\theta_{13}$.
%\item The short oscillation length $L_{23}$ can be determined directly and can be compared with the predictions of the ``global'' analysis,
\item The background from the solar neutrino events (whose total number is by a factor of two less than the expected effect) can be directly measured (by removing the source) and systematic uncertainty can be determined from the measurements with a different source of Table \ref{tab:sources}. 
\end{itemize}
The disadvantage is that some spurious $L$-dependence of the event rate comes from the geometry. This, however, can be taken care of by the geometric factor $g_{av}$.

\underline{Acknowledgments}:  JDV is indebted to Prof. Jose Valle, the PASCOS10 organizing committee and Dr M. Gomez for their hospitality during PASCOS10, while Yu. N. N.  to T. Enqvist, A. Erykalov, F.v. Feilitzsch, J. Hissa, K. Loo, J. Maalampi, D. Nesterenko, L. Oberauer, F. Thurne, W. Trzaska and M. Wurm for useful discussions and private communications.
\section{References}
%\bibliography{Tenu}
\providecommand{\newblock}{}

\end{document}